\documentclass[a4paper,12pt]{article}

\pdfoutput=1

\usepackage[tbtags]{amsmath}
\usepackage{amssymb}
\usepackage{ifthen}
\usepackage{slashed}
\usepackage{amsfonts}
\usepackage{mathrsfs}
\usepackage{bm}
\usepackage{graphicx,subfigure,booktabs}
\usepackage[numbers,sort&compress]{natbib}
\usepackage{verbatim}
\usepackage{appendix}

\usepackage[colorlinks,
            linkcolor=black,
            filecolor=black,
            anchorcolor=black,
            urlcolor=black,
            citecolor=blue
            ]{hyperref}

\usepackage{color}
\usepackage{ulem}
\usepackage{setspace}
\numberwithin{equation}{section}

\newlength{\dinwidth}
\newlength{\dinmargin}
\makeatletter
\newcommand{\thickhline}{%
    \noalign {\ifnum 0=`}\fi \hrule height 1pt
    \futurelet \reserved@a \@xhline
}
\makeatother

\allowdisplaybreaks

\begin{document}

\title{\bf Three-gluon decays of radially excited quarkonia $\psi(2S)$ and $\Upsilon(2S)$ with both relativistic and QCD radiative corrections}

\author{Chao-Jie Fan$^{a,b}$ and Jun-Kang He$^{a,b}$\footnote{hejk@hbnu.edu.cn}\\[15pt]
{$^a$\small College of Physics and Electronic Science, Hubei Normal University, Huangshi 435002, China}\\[0.2cm]
{$^b$\small Key Laboratory of Quark and Lepton Physics (MOE), }\\
{\small Central China Normal University, Wuhan 430079, China}}
\date{}

%%%%%%%%%%%%%%%%%%%%%%%%%%%%%%%%%%%%%%%%%%%%%%%%%%%%%%%%%%%%%%%%%%%%%
%\date{2026.02}
%%%%%%%%%%%%%%%%%%%%%%%%%%%%%%%%%%%%%%%%%%%%%%%%%%%%%%%%%%%%%%%%%%%%%

\maketitle
\vspace{0.2cm}

%%%%%%%%%%%%%%%%%%%%%%%%%%%%%%%%%%%%%%%%%%%%%%%%%%%%%%%%%%%%
\begin{abstract}

For the radially excited heavy quarkonia $V=\psi(2S)$ and $\Upsilon(2S)$, the nodal structure of the wave function renders the three-gluon decay $V\to ggg$ acutely sensitive to relativistic corrections, a longstanding challenge for reliable theoretical predictions. Within the Bethe-Salpeter formalism under the covariant instantaneous ansatz, we construct analytic harmonic-oscillator wave functions incorporating the $2S$ node and derive model-independent relations among the polarized decay widths from helicity-flip and phase-space symmetries. Motivated by the strikingly slow $\hat{q}^{2}$-order convergence driven by destructive interference at the node, we introduce a concise phenomenological treatment of the higher-order contributions that preserves the correct low-momentum limit. Including both relativistic and QCD radiative corrections, our predictions for $\Gamma(V\to ggg)$, $\Gamma(V\to e^{+}e^{-})$ and $R_{V}$ agree well with experiment, and the extracted $\beta_{V}$ lies at the lower end of typical phenomenological ranges, reflecting a more localized momentum-space wave function.

\end{abstract}

\maketitle

%%%%%%%%%%%%%%%%%%%%%%%%%%%%%%%%%%%%%%%%%%%%%%%%%%%%%%%%%%%%%%%%%%%%%%

\newpage

%%%%%%%%%%%%%%%%%%%%%%%%%%%%%%%%%%%%%%%%%%%%%%%%%%%%%%%%%%%%%%%%%%%%%
\section{INTRODUCTION}
\label{sec:intro}
%%%%%%%%%%%%%%%%%%%%%%%%%%%%%%%%%%%%%%%%%%%%%%%%%%%%%%%%%%%%%%%%%%%%%%

Charmonium states provide a unique laboratory for testing the interplay between perturbative and non-perturbative QCD, as well as the role of relativistic effects in heavy-quark bound systems. With $  \langle v^2 \rangle \approx 0.3  $ for charmonium, relativistic corrections of relative order $  v^2  $ are no longer negligible and must be systematically included for precision phenomenology~\cite{Bodwin:1994jh, QuarkoniumWorkingGroup:2004kpm}. The radially excited vector state $  \psi(2S)  $ is of particular interest: its larger radius and the node in the radial wave function lead to enhanced relativistic dynamics and characteristic cancellations in both production and decay amplitudes, offering a sensitive probe beyond the ground state $  J/\psi  $.

Over the past two decades, relativistic corrections for $  \psi(2S)  $ have been systematically investigated in both production and decay processes, progressively revealing their indispensable role. Studies in diffractive photoproduction at HERA revealed that the $  \sigma(\psi(2S))/\sigma(J/\psi)  $ ratio is suppressed by the radial node~\cite{H1:2002yab, ZEUS:2022sxn}. The nonrelativistic QCD (NRQCD) approach has also been applied to $  \psi(2S)  $ production, where predictions for yields and especially polarization observables for $  \psi(2S)  $ still exhibit some discrepancies with experimental data at hadron colliders~\cite{Shao:2014yta}. In the same period, the instantaneous Salpeter method was applied to annihilation decays, demonstrating that the node structures in the $  \psi(2S)  $ radial wave function generate large relativistic corrections even for bottomonium, leading to a slower fall-off of three-gluon and leptonic widths with radial quantum number~\cite{Fu:2010gd}. By the late 2010s, full Bethe-Salpeter (B-S) calculations for semileptonic decays to orbitally and radially excited charmonia (including $  \psi(2S)  $) showed that relativistic effects can exceed $50\%$ for excited states~\cite{Geng:2018qrl}. Lattice QCD determinations of B-S wave functions for $  \psi(2S)  $ further provided non-perturbative input, confirming the nodal structure~\cite{Nochi:2016wqg}.

In the 2020s, the field has advanced to high-precision NRQCD applications. For hadroproduction, leading-power fragmentation supplemented by $  v^2  $ corrections has been applied to high-$  p_T  $ prompt $  \psi(2S)  $ at the LHC. Most recently, Bertone et al.~\cite{Bertone:2025jex} performed the first dedicated study and found that $  v^2  $ corrections increase the gluon-fragmentation cross section by factors of $3.3-6.6$ (depending on $  \langle v^2 \rangle = 0.25-0.50 $) while affecting the charm channel only moderately; the resulting predictions achieve excellent agreement with ATLAS and CMS data at $  \sqrt{s}=13  $ TeV for $  p_T \gtrsim 60  $ GeV and also improve the description of polar anisotropy $  \lambda_\theta  $. Similarly, Copeland et al.~\cite{Copeland:2025osx} investigated $  \psi(2S)  $ production inside jets at LHCb using NRQCD combined with fragmenting jet functions and a Gluon Fragmentation Improved Pythia framework, demonstrating high sensitivity of the jet substructure variable $  z = p_T(\psi(2S))/p_T(\rm jet)  $ to the long-distance matrix elements and providing new constraints on the excited-state wave-function parameters. These latest works, together with earlier double-charmonium studies~\cite{He:2024ugx}, confirm that relativistic corrections and wave-function details (including implicit nodal effects through LDMEs) are indispensable for accurate $  \psi(2S)  $ production phenomenology across a wide kinematic range. On the decay side, $  v^2  $ corrections to exclusive hadronic channels $  \psi(nS) \to \rho\pi  $ and $  \gamma\pi  $ were shown to play a key role in resolving the $  \rho\pi  $ puzzle for $  \psi(2S)  $~\cite{Kivel:2023fgu}, while in the exclusive hadronic decay $  \psi(2S) \to p\bar{p}  $, the $  v^2  $ corrections with $  \langle v^2 \rangle \approx 0.64  $ dominate the amplitude, leading to their crucial role in understanding the decay process and a worsened convergence of the $v^2$ expansion~\cite{Kivel:2022qjy}. Most recently, the B-S formalism has yielded complete analytic results for vector quarkonium $ J/\psi \to ggg $, with relativistic corrections reaching $  \sim 77\%  $ and paving the way for excited-state extensions~\cite{Jiang:2025cks}.

The radially excited quarkonium decay $ V \to ggg  $ (with $V = \psi(2S)$ or $\Upsilon(2S)$) stands out as an ideal channel. The purely gluonic final state eliminates hadronization uncertainties, allowing the nodal structure of the long-distance B-S wave function to be probed through its convolution with the hard-scattering amplitude. This process is therefore well suited to probe $  v^2  $-order relativistic corrections originating primarily from the initial-state quarkonium structure. Although the destructive interference induced by the radial node has been repeatedly highlighted in production studies, it has not yet been isolated using analytic B-S wave functions in the decay $ V \to ggg  $. A relativistic B-S calculation incorporating the node in this channel, with all eight Dirac structures of the covariant amplitude kept consistently, is therefore still missing, posing an ongoing challenge in providing reliable and precise predictions.

In this work we compute the complete $  q^2  $-order relativistic corrections to the three-gluon decays $ \psi(2S) \to ggg  $ and $ \Upsilon(2S) \to ggg  $, by employing the analytic form of the B-S wave function for the radially excited states $ \psi(2S) $ and $\Upsilon(2S)$ given in our latest work~\cite{Jiang:2025cks}. We find that the $  q^2  $-order relativistic corrections in the $  \psi(2S) \to ggg  $ process are significantly larger than those in the $  J/\psi \to ggg  $ process, with the correction factor $  \kappa \beta^2_{V} / M^2  $ approximately equal to $0.77$ for $  J/\psi  $ and $2.7$ for $  \psi(2S)  $, resulting in an unphysical negative  decay width for the $  \psi(2S)  $ channel. To obtain a physically meaningful prediction, we adopt a phenomenological treatment of the hard kernel in the $ V \to ggg  $ process, as proposed in Ref.~\cite{Chao:1995cz}, which effectively incorporates higher-order (beyond $  q^2  $-order) relativistic corrections and the nodal structure effects of the radially excited state $ V $. And our theoretical predictions for the branching ratios $\mathcal{B}(V \to ggg)$ and $\mathcal{B}(V \to e^+ e^-)$ are consistent with experimental data. Moreover, we extract the harmonic oscillator parameter  $\beta_{V}$, which provides valuable input for calibrating non-perturbative models of quarkonium structure.

The paper is organized as follows. The theoretical framework and the analytic calculation for the decay process $V \to  ggg$ are presented in detail in Sec.~\ref{sec:framework}. In Sec.~\ref{sec:Results and discussions} we show our results and phenomenological discussions. The last section is our summary.

%%%%%%%%%%%%%%%%%%%%%%%%%%%%%%%%%%%%%%%%%%%%%%%%%%%%%%%%%%%%%%%%%%%%%
\section{THEORETICAL FRAMEWORK}
\label{sec:framework}
%%%%%%%%%%%%%%%%%%%%%%%%%%%%%%%%%%%%%%%%%%%%%%%%%%%%%%%%%%%%%%%%%%%%%
\subsection{Bethe-Salpeter equation}
\label{subsec:B-S equation}
%%%%%%%%%%%%%%%%%%%%%%%%%%%%%%%%%%%%%%%%%%%%%%%%%%%%%%%%%%%%%%%%%%%%%

The B-S equation~\cite{Salpeter:1951sz,Salpeter:1952ib} provides a covariant, relativistic framework for describing bound states. In this work it is implemented under the covariant instantaneous ansatz (CIA)~\cite{Mitra:1990av,Negash:2015rua,Bhatnagar:2016otj}, in which the interaction kernel is reduced to an instantaneous potential while the B-S amplitude is kept fully covariant. This implementation is particularly well-suited to the systematic treatment of radially excited quarkonia, where the nodal structure of the wave function is crucial. In this formalism, the B-S wave function $\Psi(K, q)$ encodes the complete bound-state information and is determined by the dynamical equation~\cite{Mengesha:2011pu, Negash:2015hma, Negash:2015rua, Bhatnagar:2016otj, Gebrehana:2019mpw, He:2020kin, Jiang:2025cks}
\begin{equation}\label{bse}
S^{-1}_{F}(f)\Psi(K,q)S^{-1}_{F}(-\bar{f}) =
\int\frac{\mathrm{d}^{4}q^{\prime}}{(2\pi)^{4}}\Big{[}-i\mathcal{K}(K,q,q^{\prime})\Psi(K,q^{\prime})\Big{]}.
\end{equation}
Applying this formalism to heavy quarkonium $V$ (with $V = \psi(2S)$ or $\Upsilon(2S)$), $\mathcal{K}(K,q,q^{\prime})$ represents the interaction kernel between the internal quark and antiquark, and $S_{F}(p)=i/(\slashed{p}-\hat{m}+i\epsilon)$ is the propagator with the effective mass $\hat{m}$ of the $c$ or $b$ quark. The momenta of the quark and antiquark are given by
\begin{equation}
f=\frac{K}{2}+q, \qquad \bar{f}=\frac{K}{2}-q,
\end{equation}
where $q$ and $K$ represent the internal momentum and the total momentum of the quarkonium, respectively.

For convenience,  the internal momentum $q$ is decomposed into a transverse component $\hat{q}$ (with $\hat{q} \cdot K = 0$) and a longitudinal component $q_{\parallel}$ (parallel to the total momentum $K$):
\begin{eqnarray}
q^{\mu}&=& q_{\parallel}^{\mu}+\hat{q}^{\mu}, \quad\quad \quad\quad q_{\parallel}^{\mu} = \frac{q_{K}}{M} K^{\mu}.
\end{eqnarray}
Here both $q_{K}=q\cdot K / M$ and $\hat{q}^{2}=q^{2}-q_{K}^{2}$ are Lorentz invariant variables and $M$ is the mass of the heavy  quarkonium.

The CIA~\cite{Mitra:1990av, Bhatnagar:2009jg, Bhatnagar:2013bha, Bhatnagar:2016otj, Gebrehana:2019mpw} posits that the interaction is instantaneous, so that the interaction kernel $\mathcal{K}(K,q,q^{\prime})$ is independent of the timelike component of the internal momentum and reduces to a function of the transverse relative momenta only. In this framework the B-S kernel is taken, in the spin-color sector, to be of one-gluon-exchange type~\cite{Negash:2015rua,Bhatnagar:2016otj},
\begin{equation}\label{kernelOGE}
\mathcal{K}(K,q,q^{\prime}) = \Bigl(\tfrac{1}{2}\vec{\lambda}_{1}\!\cdot\!\tfrac{1}{2}\vec{\lambda}_{2}\Bigr)(\gamma_{\mu}\otimes\gamma^{\mu})\, V(\hat{q}-\hat{q}^{\prime}),
\end{equation}
while its scalar spatial part $V(\hat{q},\hat{q}^{\prime})$ contains, in a single structure, a long-distance harmonic confinement that smoothly interpolates towards a linear confinement for heavy quarkonia~\cite{Negash:2015rua,Mitra:1990av}:
\begin{eqnarray}\label{Vspatial}
V(\hat{q},\hat{q}^{\prime}) &=& \frac{3}{4}\,\omega_{q\bar{q}}^{2}\!\int \mathrm{d}^{3}\vec{r}\,
\Bigl[\,r^{2}\,\bigl(1+4\hat{m}_{1}\hat{m}_{2}A_{0}M_{>}^{2}r^{2}\bigr)^{-1/2} - \tfrac{C_{0}}{\omega_{0}^{2}}\,\Bigr]\,
e^{i(\hat{q}-\hat{q}^{\prime})\cdot\vec{r}},\nonumber\\
\omega_{q\bar{q}}^{2} &=& 4\,M_{>}\,\hat{m}_{1}\hat{m}_{2}\,\omega_{0}^{2}\,\alpha_{s}(M_{>}^{2}),\qquad M_{>}=\max(M,m_{1}+m_{2}),
\end{eqnarray}
where $\hat{m}_{i}=m_{i}/(m_{1}+m_{2})$, $\omega_{0}$ is the basic harmonic spring constant, $C_{0}/\omega_{0}^{2}$ is a constant that takes account of the correct zero-point energies, and $A_{0}$ controls the harmonic--to--linear interpolation. The running of $\omega_{q\bar{q}}^{2}$ through $\alpha_{s}(M_{>}^{2})$ encodes the asymptotic-freedom dependence of the short-range interaction, while the polynomial-in-$r$ structure encodes confinement. In the heavy-quarkonium limit ($\omega\simeq m$) this kernel can be cast, in momentum space, in the compact local form~\cite{Negash:2015rua}
\begin{equation}\label{Vlocal}
V(\hat{q},\hat{q}^{\prime}) = \bar{V}(\hat{q})\,\delta^{3}(\hat{q}-\hat{q}^{\prime}),\qquad
\bar{V}(\hat{q}) = \omega_{q\bar{q}}^{2}\Bigl[\,\kappa\,\vec{\nabla}_{\hat{q}}^{\,2} + \tfrac{C_{0}}{\omega_{0}^{2}}\,\Bigr](2\pi)^{3},
\end{equation}
so that the coupled Salpeter equations for the scalar coefficients of the B-S amplitude reduce to an approximate three-dimensional harmonic-oscillator equation; the harmonic-oscillator form in Eq.~(\ref{fhatq}) below is the corresponding $2S$ analytic solution, with the parameter $\beta_{V}$ determined in Sec.~\ref{sec:Results and discussions} from the experimental ratio $R_{V}$. The corresponding Salpeter wave function is
\begin{equation}
\psi(\hat{q})=\frac{i}{2\pi}\int\mathrm{d}q_{K}\Psi(K,q).
\end{equation}

For a vector quarkonium $  V  $, the complete decomposition of the Salpeter wave function in terms of various Dirac structures is~\cite{LlewellynSmith:1969az, Wang:2005qx, Bhatnagar:2013bha, Negash:2015rua, Jiang:2025cks, Alkofer:2002bp},
\begin{eqnarray}\label{psihatq}
\psi(\hat{q})&=&M\slashed{\varepsilon} f_{1}(\hat{q})+\slashed{\varepsilon}\slashed{K} f_{2}(\hat{q})
+[\slashed{\varepsilon}\slashed{\hat{q}}-\hat{q}\cdot \varepsilon]f_{3}(\hat{q}) +[\slashed{K}\slashed{\varepsilon}\slashed{\hat{q}}-(\hat{q}\cdot\varepsilon)\slashed{K}]
\frac{f_{4}(\hat{q})}{M}\nonumber\\
& &+(\hat{q}\cdot\varepsilon)f_{5}(\hat{q})+(\hat{q}\cdot\varepsilon) \slashed{K}\frac{f_{6}(\hat{q})}{M} + (\hat{q}\cdot\varepsilon)\slashed{\hat{q}}f_{7}(\hat{q}) + (\hat{q}\cdot\varepsilon) \slashed{K} \slashed{\hat{q}} \frac{f_{8}(\hat{q})}{M},
\end{eqnarray}
where each scalar function $f_{i}(\hat{q})$ ($i=1,2,\cdots,8$) depends on $\hat{q}^{2}$, and $\varepsilon$ is the polarization vector of $V$. Equation~(\ref{psihatq}) is the most general Lorentz-covariant decomposition of the Salpeter amplitude allowed by the $J^{PC}=1^{--}$ quantum numbers~\cite{LlewellynSmith:1969az,Bhatnagar:2013bha,Negash:2015rua}. Within the CIA framework of Refs.~\cite{Mitra:1990av,Negash:2015rua,Bhatnagar:2016otj}, the eight scalar coefficients $f_{i}(\hat{q})$ satisfy a coupled set of Salpeter integral equations whose interaction kernel is the one specified in Eqs.~(\ref{kernelOGE})--(\ref{Vlocal}). In the heavy-quarkonium limit those coupled equations reduce to an approximate three-dimensional harmonic-oscillator equation, and the $f_{i}(\hat{q})$ become proportional to a single common radial function $f(\hat{q})$ of harmonic-oscillator form, with calculable relative weights ordered by the power-counting rule of Refs.~\cite{Bhatnagar:2005vw,Bhatnagar:2013bha} in powers of $\hat{q}/M$. In the present work the coupled Salpeter equations are not solved numerically; we instead take over this analytic solution and, to capture the radial node of the $2S$ states, supplement it with the standard first-radial-excitation factor $\bigl(1-2\hat{q}^{2}/3\beta_{V}^{2}\bigr)$ in the oscillator basis. The only phenomenological input left is the harmonic-oscillator parameter $\beta_{V}$, fixed in Sec.~\ref{sec:Results and discussions} from the experimental ratio $R_{V}$. In our recent study of ground-state decays~\cite{Jiang:2025cks} only the Dirac structures up to subleading order, i.e., ${\cal O}(\hat{q})$, were retained. In this work, we extend the framework by including all independent structures, i.e., those up to ${\cal O}(\hat{q}^{2})$. Following Refs.~\cite{Bhatnagar:2005vw,Bhatnagar:2013bha}, all eight scalar coefficients $f_{i}(\hat{q})$ in Eq.~(\ref{psihatq}) are kept and approximated in terms of one common scalar radial function $f(\hat{q})$ with relative weights fixed by the power-counting rule up to ${\cal O}(\hat{q}^{2}/M^{2})$ and by matching to the NRQCD spin projector in the limit $m\simeq M/2$~\cite{Bodwin:2002cfe}. This procedure provides an approximate closed form of the complete Salpeter amplitude that is consistent with the order at which the $\hat{q}^{2}$-order relativistic corrections to the decay amplitude are computed; it is not a truncation that drops Dirac components. The resulting compact expression for $\psi(\hat{q})$ reads:
\begin{eqnarray}\label{psi_simple}
\psi(\hat{q})&=& \left(M\slashed{\varepsilon} - \slashed{\varepsilon}\slashed{K} + 2 (\hat{q}\cdot \varepsilon) - \frac{2 \slashed{K}(\hat{q}\cdot \varepsilon)}{M} + \frac{2 \slashed{K}\slashed{\varepsilon}\slashed{\hat{q}} }{M} + \frac{2 (\hat{q}\cdot \varepsilon)\slashed{\hat{q}} }{M} - \frac{2 (\hat{q}\cdot \varepsilon)\slashed{K}\slashed{\hat{q}} }{M^{2}}\right) f(\hat{q}),
\end{eqnarray}
where the Dirac structures of the wave function are consistent with the spin projector commonly adopted in the NRQCD approach with the limit $  m \approx M/2  $~\cite{Bodwin:2002cfe}.
The scalar function is taken as
\begin{eqnarray}\label{fhatq}
f(\hat{q})&=&N_{V}(\frac{3}{2})^{1/2}\frac{1}{\pi^{3/4}}\frac{1}{\beta^{3/2}_{V}} \left(1-\frac{2 \mathbf{\hat{q}}^{2} }{3 \beta^{2}_{V}}\right) e^{-\frac{\mathbf{\hat{q}}^{2}}{2\beta^{2}_{V}}},
\end{eqnarray}
where $  N_{V}  $ is a constant and $  \beta_{V}  $ is the harmonic oscillator parameter. To retain the radial node structure of the excited state $V$, we adopt the above harmonic-oscillator form for the scalar function $f(\hat{q})$. We stress that within the CIA framework of Refs.~\cite{Mitra:1990av,Negash:2015rua,Bhatnagar:2016otj} this is not an ad hoc ansatz competing with the OGE-type kernel: the spatial part $V(\hat{q},\hat{q}^{\prime})$ specified in Eqs.~(\ref{kernelOGE})--(\ref{Vlocal}) is constructed so that, in the heavy-quark limit, the coupled Salpeter equations for the scalar coefficients of Eq.~(\ref{psihatq}) reduce to a three-dimensional harmonic-oscillator equation~\cite{Negash:2015rua}; Eq.~(\ref{fhatq}) is the corresponding $2S$ analytic solution and the nodal factor $(1-2\hat{q}^{2}/3\beta_{V}^{2})$ is its first radial excitation. The instantaneous reduction of the kernel neglects the retardation of the exchanged gluon, whose leading effect is of order $v^{2}$ relative to the static potential and generates Breit--Fermi-type terms (spin--orbit, tensor, Darwin) in the bound-state equation. Such retardation and Breit--Fermi corrections enter our calculation of $V\to ggg$ only through the bound-state wave function, and their net effect is absorbed into the fitted value of $\beta_{V}$. The $v^{2}$ relativistic corrections that enter the decay amplitude dynamically -- through the small Dirac components of Eq.~(\ref{psihatq}) and through the expansion of the hard kernel in $\hat{q}$ -- are kept explicitly, and are precisely those that are resummed to all orders in the phenomenological improvement introduced later in Sec.~\ref{subsec:V-ggg} following Ref.~\cite{Chao:1995cz}.

This extended form of $\psi(\hat{q})$ systematically includes relativistic spin-orbit couplings, providing a consistent and improved parametrization of the Salpeter wave function. For convenience, we write it as $\psi(\hat{q})={\cal P}(\hat{q})\,f(\hat{q})$, with ${\cal P}(\hat{q})$ denoting the Dirac-structure factor in parentheses of Eq.~(\ref{psi_simple}).

%%%%%%%%%%%%%%%%%%%%%%%%%%%%%%%%%%%%%%%%%%%%%%%%%%%%%%%%%%%%%%%%%%%%%%%%%%%%%%%%%%%%%%%%%%%%%%%%%%%%%%%%%%%
\subsection{The decay $V  \to  ggg$ in perturbative QCD}
\label{subsec:V-ggg}
%%%%%%%%%%%%%%%%%%%%%%%%%%%%%%%%%%%%%%%%%%%%%%%%%%%%%%%%%%%%%%%%%%%%%%%%%%%%%%%%%%%%%%%%%%%%%%%%%%%%%%%%%%%

Within the B-S formalism, the decay amplitude for $V \to ggg$ is obtained through the convolution of the bound-state wave function with a perturbative hard-scattering kernel. Since the wave function $\Psi(K,q)$ encodes the full information about the relativistic internal motion, this framework allows a systematic treatment of the higher-order corrections induced by the nodal structure. In the rest frame of $V$, the amplitude can be written explicitly as
\begin{eqnarray}
{\mathcal A}&=&\sqrt{N_{c}}\int\frac{\mathrm{d}^{4}q}{(2\pi)^{4}}\textrm{Tr}\left[\Psi(K,q){\cal O}(f,\bar{f})\right],
\end{eqnarray}
where the factor $\sqrt{N_{c}}$ reflects the color-singlet nature of the quark-antiquark pair, ${\cal O}(f,\bar{f})$ denotes the hard-scattering amplitude for $q\bar{q} \to ggg$, and the momenta of the constituent quark and antiquark are
\begin{eqnarray}
f^{\mu}&=&\frac{K^{\mu}}{2}+q^{\mu}
=\left(\frac{M}{2}+q^{0},\mathbf{q}\right),\nonumber\\
\bar{f}^{\mu}&=&\frac{K^{\mu}}{2}-q^{\mu}
=\left(\frac{M}{2}-q^{0},-\mathbf{q}\right).
\end{eqnarray}
Under the CIA~\cite{Mitra:1990av, Bhatnagar:2009jg, Bhatnagar:2013bha, Bhatnagar:2016otj, Gebrehana:2019mpw}, it is consistent to treat $q^{0}\ll M$. Under this approximation, the momenta reduce to
\begin{eqnarray}
f^{\mu}&\approx&\left(\frac{M}{2},\mathbf{q}\right)=\frac{K^{\mu}}{2}+\hat{q}^{\mu},\nonumber\\
\bar{f}^{\mu}&\approx&\left(\frac{M}{2},-\mathbf{q}\right)=\frac{K^{\mu}}{2}-\hat{q}^{\mu},
\end{eqnarray}
and the hard-scattering amplitude simplifies to
\begin{eqnarray}
{\cal O}(f,\bar{f})&\approx&{\cal O}(\hat{q}).
\end{eqnarray}
From another point of view~\cite{Chao:1995cz}, this treatment can be connected with the on-shell condition, which maintains the gauge invariance of the hard-scattering amplitude.

With the above approximations, the decay amplitude of $V \to ggg$ takes the compact form
\begin{eqnarray}
{\mathcal A}&=&-i\sqrt{N_{c}}\int\frac{\mathrm{d}^{3}\hat{q}}{(2\pi)^{3}} f(\hat{q}) \textrm{Tr}\left[{\cal P}(\hat{q}){\cal O}(\hat{q})\right].
\end{eqnarray}
The explicit expression for the hard-scattering amplitude ${\cal O}(\hat{q})$ is
\begin{eqnarray}
{\cal O}(\hat{q}) & = & \frac{-ig^{3}_{s}}{4 N_{c}}
\bigg{(} \textrm{Tr} [T^{a_{1}}T^{a_{2}}T^{a_{3}}]
I(k_{1},k_{2},k_{3},\epsilon_{1},\epsilon_{2},\epsilon_{3})
+ \textrm{Tr} [T^{a_{2}}T^{a_{1}}T^{a_{3}}]
I(k_{2},k_{1},k_{3},\epsilon_{2},\epsilon_{1},\epsilon_{3}) \nonumber\\
& + & \textrm{Tr} [T^{a_{2}}T^{a_{3}}T^{a_{1}}]
I(k_{2},k_{3},k_{1},\epsilon_{2},\epsilon_{3},\epsilon_{1})
+ \textrm{Tr}[T^{a_{3}}T^{a_{2}}T^{a_{1}}] I(k_{3},k_{2},k_{1},\epsilon_{3},\epsilon_{2},\epsilon_{1})\nonumber\\
& + &\textrm{Tr}[T^{a_{3}}T^{a_{1}}T^{a_{2}}]
I(k_{3},k_{1},k_{2},\epsilon_{3},\epsilon_{1},\epsilon_{2})
+ \textrm{Tr}[T^{a_{1}}T^{a_{3}}T^{a_{2}}]
I(k_{1},k_{3},k_{2},\epsilon_{1},\epsilon_{3},\epsilon_{2}) \bigg{)}
\end{eqnarray}
with
\begin{eqnarray}
I&&(k_{1},k_{2},k_{3},\epsilon_{1},\epsilon_{2},\epsilon_{3}) \nonumber\\
& & = \frac{ \slashed{\epsilon}^{*}_{1} \cdot (\slashed{k}_1 - \slashed{k}_2 - \slashed{k}_3 + 2 \hat{\slashed{q}} + M )\cdot \slashed{\epsilon}^{*}_{2} \cdot( \slashed{k}_1 + \slashed{k}_2 - \slashed{k}_3 + 2 \hat{\slashed{q}} + M )\cdot \slashed{\epsilon}^{*}_{3}}{[ -k_1 \cdot k_2 - k_1 \cdot k_3 + 2(k_1 \cdot \hat{q}) + \hat{q}^2 + i \varepsilon][-k_1 \cdot k_3 - k_2 \cdot k_3 - 2(k_3 \cdot \hat{q}) + \hat{q}^2 + i \varepsilon]} ,
\end{eqnarray}
where $k_{i}$ and $\epsilon_{i}$ ($i=1,2,3$) are the momenta and polarization vectors of the outgoing gluons. Using the color identity $\textrm{Tr}[T^{a_{1}}T^{a_{2}}T^{a_{3}}]=(d^{a_{1}a_{2}a_{3}}+if^{a_{1}a_{2}a_{3}})/4$, the amplitude decomposes into a part proportional to the symmetric structure constant $d^{a_{1}a_{2}a_{3}}$ and another proportional to the antisymmetric constant $f^{a_{1}a_{2}a_{3}}$. The latter contribution vanishes identically after integration over the internal momentum $\hat{q}$. Thus the hard-scattering amplitude reduces to
\begin{eqnarray}
{\cal O}(\hat{q}) & = & \frac{-ig^{3}_{s}d^{a_{1}a_{2}a_{3}}}{16 N_{c}}
\bigg{(} I(k_{1},k_{2},k_{3},\epsilon_{1},\epsilon_{2},\epsilon_{3})
+ \textrm{ 5 permutations of 1, 2 and 3.}  \bigg{)}.
\end{eqnarray}

The decay width is obtained from
\begin{eqnarray}\label{gamma}
\Gamma^{ggg}& = & \frac{1}{2 M}\int \frac{1}{3} \frac{1}{3!} \sum_{\mathrm{color, pol.}} \mid {\mathcal A} \mid^{2} \mathrm{d}\Phi_{3},
\end{eqnarray}
where $\int \mathrm{d}\Phi_{3}$ denotes the integration over the three-body phase space,  the factor $1/3$ averages over the initial quarkonium polarizations, $1/3!$ accounts for the identical gluons in the final state, and the sum runs over the colors and polarizations of all external particles.

The decay $V \to ggg$ possesses a soft-gluon singularity, which gives rise to a potential infrared (IR) divergence. When the full $\hat{q}$-dependence of the hard amplitude is retained, we find the integration over the internal momentum $\hat{q}$ becomes divergent starting at ${\cal O}(\hat{q}^4)$. Physically, this divergence corresponds to the process in which a soft gluon converts the incoming $S$-wave color-singlet quarkonium into a $P$-wave color-octet intermediate state~\cite{Bodwin:2002cfe}. Within the established NRQCD factorization formalism, such IR divergences can be systematically absorbed into the non-perturbative matrix elements of $P$-wave color-octet operators~\cite{Bodwin:2002cfe}. Phenomenological determinations of these higher-order matrix elements remain uncertain, but power-counting arguments suggest that ${\cal O}(v^4)$ contributions are generally suppressed compared with the leading ${\cal O}(v^2)$ terms~\cite{Bodwin:2002cfe, Jiang:2025cks}.

Up to $\hat{q}^2$-order relativistic corrections, the modulus squared of the decay amplitude is obtained as
\begin{eqnarray}\label{squaredamplitude}
\mid {\mathcal A} \mid^{2} & = & \left| \sqrt{N_{c}} \int\frac{\mathrm{d}^{3}\hat{q}}{(2\pi)^{3}} f(\hat{q}) {\cal M}^{0}(0) \right|^{2} \nonumber\\
& + & 2 \textrm{Re} \left[ N_{c} \int \frac{\mathrm{d}^{3}\hat{q}}{(2\pi)^{3}} \frac{\mathrm{d}^{3} \hat{q}^{\prime}}{(2\pi)^{3}} f(\hat{q}) f^{*}(\hat{q}^{\prime})  {\cal M}^{0*}(0)  {\cal M}^{1}(0) \right]
\end{eqnarray}
with
\begin{eqnarray}
{\cal M}^{0}(0) & = &\textrm{Tr}\left[ {\cal P}(0){\cal O}(0)\right]
\end{eqnarray}
and
\begin{eqnarray}
{\cal M}^{1}(0) & = & \frac{\hat{q}^{\alpha}\hat{q}^{\beta} }{2!} \left[ \left. \frac{\partial \textrm{Tr} \left[{\cal P}(\hat{q}){\cal O}(\hat{q})\right]}{\partial \hat{q}^{\alpha} \partial \hat{q}^{\beta}} \right|_{\hat{q}=0}\right].
\end{eqnarray}
Here we expand the modulus squared $\mid {\mathcal A} \mid^{2}$ to quadratic order in the internal momentum $\hat{q}$, and make the following substitution to project out the $S$-wave state:
\begin{eqnarray}
\hat{q}^{\alpha}\hat{q}^{\beta} &  \to  & \frac{\mathbf{\hat{q}}^{2}}{3}\left( -g^{\alpha\beta} + \frac{K^{\alpha}K^{\beta}}{M^{2}} \right),
\end{eqnarray}
so that ${\cal M}^{1}(0) $ becomes
\begin{eqnarray}
{\cal M}^{1}(0) & = &\frac{1}{2!} \frac{\mathbf{\hat{q}}^{2} }{3}  \left.\left[ \left( -g^{\alpha\beta} + \frac{K^{\alpha}K^{\beta}}{M^{2}} \right) \frac{\partial \textrm{Tr} \left[{\cal P}(\hat{q}){\cal O}(\hat{q}) \right]}{\partial \hat{q}^{\alpha} \partial \hat{q}^{\beta}} \right|_{\hat{q}=0} \right].
\end{eqnarray}

Performing the integration over the three-body phase space in Eq.~(\ref{gamma}) yields the decay width including $  \hat{q}^2  $-order relativistic corrections:
\begin{eqnarray}\label{GammaResult1}
\Gamma^{ggg}& = & \frac{10 (\pi^2 - 9 ) \alpha_s^3}{81 \pi^{6} M} \int \mathrm{d}^{3}\hat{q} \mathrm{d}^{3}\hat{q}^{\prime} f(\hat{q}) f(\hat{q}^{\prime}) \left(1-\frac{48+121\pi^2}{48(\pi^2 - 9)}\frac{\mathbf{\hat{q}}^2}{M^2}\right).
\end{eqnarray}
However, our subsequent numerical evaluation for the radially excited state $  \psi(2S)  $ (presented later in this work) reveals that this expression produces an unphysical negative decay width. Inspired by the phenomenological improvement proposed in Ref.~\cite{Chao:1995cz}, we therefore introduce the following improved expression:
\begin{eqnarray}\label{GammaResult2}
\Gamma^{ggg}& = & \frac{10 (\pi^2 - 9 ) \alpha_s^3}{81 \pi^{6} M} \left| \int \mathrm{d}^{3}\hat{q} f(\hat{q}) \left[1+\frac{48+121\pi^2}{96(\pi^2 - 9)}\frac{\mathbf{\hat{q}}^2}{M^2}\right]^{-1} \right|^{2}.
\end{eqnarray}
This form preserves the $  \hat{q}^2  $-order relativistic correction in the small-$  \hat{q}  $ limit while effectively incorporating partial higher-order contributions at large momentum. It guarantees a positive definite decay width for $  \psi(2S)  $ and ensures better consistency with the full $  \hat{q}  $-dependent hard kernel across both non-relativistic and highly relativistic regimes. This improved treatment significantly enhances the convergence of the integral of the hard kernel convolved with the scalar wave function $f(\hat{q})$ across both low- and high-momentum regions: it exactly reproduces the $\hat{q}^2$-order result at small $\hat{q}$ (owing to the rapid convergence of the Taylor expansion), while the denominator provides natural suppression at large $\hat{q}$ and prevents any pathological behavior from the wave-function tail. It is worth noting that a strict perturbative expansion to $\mathcal{O}(\hat{q}^4)$ would induce an IR divergence. Although within the NRQCD factorization formalism such divergences can be absorbed into non-perturbative $P$-wave color-octet matrix elements, the phenomenological determination of these additional parameters introduces significant uncertainty for the decays of radially excited states. In this context, while our phenomenological treatment is not a first-principles derivation, its remarkably concise and physically transparent form offers a practical yet intuitive way to incorporate the dominant higher-order relativistic corrections beyond $  \hat{q}^2  $, together with the nodal structure of the radial wave function that is characteristic of the $2S$ state.

We can now exploit these expressions to study the polarization characteristics and symmetries of the decay. To discuss the symmetry in the decay process $V  \to  ggg$, the corresponding polarized decay widths are given by
\begin{eqnarray}\label{polarized widths}
\Gamma_{[\lambda_V, \lambda_1, \lambda_2, \lambda_3]}^{ggg}& = & \frac{1}{2 M}\int \frac{1}{3!} \sum_{\mathrm{color}} \mid {\mathcal A} \mid^{2} \mathrm{d}\Phi_{3}.
\end{eqnarray}
Here, $\lambda_V$ denotes the helicity of the heavy quarkonium, while $\lambda_i$ ($i=1,2,3$) represents the gluon helicity. We adopt the following conventions
\begin{eqnarray}
K & = & M (1, 0, 0, 0), \quad \varepsilon(\pm 1) = \frac{1}{\sqrt{2}}(0, \mp 1, - i, 0), \quad \varepsilon(0) = (0, 0, 0, 1)  \nonumber \\
k_{1} & = & \frac{M x_{1}}{2}(1, 0, 0, 1), \quad \epsilon_{1}(\pm 1) = \frac{1}{\sqrt{2}}(0, \mp 1, - i, 0),  \nonumber \\
k_{2} & = & \frac{M x_{2}}{2} (1, \sin\theta_{12}, 0, \cos\theta_{12}), \quad \epsilon_{2}(\pm 1) = \frac{1}{\sqrt{2}}(0, \mp \cos\theta_{12}, - i, \pm \sin\theta_{12}), \nonumber \\
k_{3} & = & \frac{M x_{3}}{2} (1, -\sin\theta_{13}, 0, \cos\theta_{13}), \quad  \epsilon_{3}(\pm 1) = \frac{1}{\sqrt{2}}(0, \mp \cos\theta_{13}, - i, \mp \sin\theta_{13}),
\end{eqnarray}
where $\theta_{ij}$ is the angle between gluons $i$ and $j$.

%%%%%%%%%%%%%%%%%%%%%%%%%%%%%%%%%%%%%%%%%%%%%%%%%%%%%%%%%%%%%%%%%%%%%%%%%%%%%%%%%%%%%%%%%%%%%%%%%%%%%%%%%%%%%%%%%%%%%%%%%%%%%%%%%
\section{RESULTS AND DISCUSSIONS}
\label{sec:Results and discussions}
%%%%%%%%%%%%%%%%%%%%%%%%%%%%%%%%%%%%%%%%%%%%%%%%%%%%%%%%%%%%%%%%%%%%%%%%%%%%%%%%%%%%%%%%%%%%%%%%%%%%%%%%%%%%%%%%%%%%%%%%%%%%%%%%%
\subsection{Polarized decay widths}
\label{subsec:Polarized decay widths}
%%%%%%%%%%%%%%%%%%%%%%%%%%%%%%%%%%%%%%%%%%%%%%%%%%%%%%%%%%%%%%%%%%%%%%%%%%%%%%%%%%%%%%%%%%%%%%%%%%%%%%%%%%%%%%%%%%%%%%%%%%%%%%%%%

The set-up of polarized decay widths $\Gamma^{ggg}_{[\lambda_V,\lambda_1,\lambda_2,\lambda_3]}$, the kinematic conventions for the helicity vectors and the gluon momenta, the helicity-flip and gluon-permutation symmetries, the resulting equivalence classes $\Gamma_1$--$\Gamma_4$ of non-vanishing configurations, and the analytic $\hat{q}^{2}$-order expressions for these polarized widths are obtained exactly as in our previous work on the ground-state decays $J/\psi\to ggg$ and $\Upsilon(1S)\to ggg$~\cite{Jiang:2025cks}, now with the Dirac structures extended to ${\cal O}(\hat{q}^{2})$ and the $1S$ scalar radial function replaced by the $2S$ one in Eq.~(\ref{fhatq}). To avoid a lengthy repetition, we summarize here only the symmetry-derived results that are directly needed for the present discussion, and collect the corresponding explicit expressions in the Appendix.

In particular, the non-vanishing polarized decay widths fall into four equivalence classes $\Gamma_1$, $\Gamma_2$, $\Gamma_3$ and $\Gamma_4$ defined in Eqs.~(\ref{group one})--(\ref{group four}) of the Appendix: $\Gamma_{1,2,3}$ each collect four equivalent helicity configurations, while $\Gamma_{4}$ collects only two. The corresponding analytic $\hat{q}^{2}$-order expressions are given in Eqs.~(\ref{gamma one first})--(\ref{gamma four first}) of the Appendix. Employing the phenomenological treatment of higher-order relativistic corrections introduced in Sec.~\ref{subsec:V-ggg} leads to the improved expressions:
\begin{eqnarray}\label{gamma one}
\Gamma_1 &=& \frac{10 \sqrt{3}  M^3 N_{V}^2 ( 3\pi^2 - 16 )^3 \alpha_s^3 }{81 \beta_{V} \pi^{9/2} (1264 + 111\pi^2)^2} \left( 1 - \frac{3 \sqrt{6\pi a_{1}} (1 + 8 a_{1})}{(1 + 24 a_{1})} e^{6 a_{1}} \operatorname{erfc}(\sqrt{6 a_{1}}) \right)^2,
\end{eqnarray}
\begin{eqnarray}\label{gamma two}
\Gamma_2 &=& \frac{10  \sqrt{3} M^3 N_{V}^2 \left(29\pi^2 -280 \right)^3 \alpha_s^3 }{81 \beta_{V} \pi^{9/2} \left(265\pi^2 -1424 \right)^2} \left(1 - \frac{3 \sqrt{6 \pi a_{2}} (1 + 8a_{2})}{(1 + 24a_{2})} e^{6a_{2}} \operatorname{erfc}\left(\sqrt{6 a_{2}}\right) \right)^2,
\end{eqnarray}
\begin{eqnarray}\label{gamma three}
\Gamma_3 &=& \frac{320 \sqrt{3} M^9 N_{V}^2 \alpha_s^3 }{81 \beta_{V}^7 \pi^{9/2} (88 + 27\pi^2)^5}\left( 1 - \frac{6 \sqrt{3 \pi} (16 + a_{3})}{\sqrt{a_{3}} (48 + a_{3})} e^{12/a_{3}} \operatorname{erfc}\left(\frac{2\sqrt{3}}{\sqrt{a_{3}}}\right) \right)^2,
\end{eqnarray}
\begin{eqnarray}\label{gamma four}
\Gamma_4 &=& \frac{20480 \sqrt{3} M^3 N_{V}^2 \left(\pi^2 - 9 \right)^3 \alpha_s^3 }{81 \beta_{V} \pi^{9/2} \left(48 + 121\pi^2\right)^2} \left( 1 - \frac{12 \sqrt{3 \pi a_{4}} (1 + 64a_{4})}{(1 + 192a_{4})} e^{48a_{4}} \operatorname{erfc}\left(4\sqrt{3 a_{4}}\right) \right)^2,
\end{eqnarray}
where $\operatorname{erfc}(x)$ is the complementary error function, and the auxiliary parameters are defined as $a_1=\frac{(3\pi^2 - 16) M^2}{(1264 + 111\pi^2) \beta_V^2}$, $a_2=\frac{(29\pi^2 - 280) M^2}{(265\pi^2 - 1424) \beta_V^2}$, $a_3=\frac{(88 + 27\pi^2) \beta_V^2}{M^2}$, and $a_4=\frac{(\pi^2 - 9) M^2}{(48 + 121\pi^2) \beta_V^2}$.

The explicit results for $\Gamma_1$ to $\Gamma_4$ are found to automatically satisfy the model-independent symmetry relations, thereby providing a direct and nontrivial verification of our calculational framework. Moreover, our numerical analysis will show that the $\hat{q}^2$-order relativistic corrections yield unphysical results for $\psi(2S)$, while the phenomenologically improved expressions give physically consistent predictions. A detailed discussion follows in the next section.

Beyond this internal consistency, these polarized decay widths offer a richer set of observables compared to the unpolarized decay width alone. They encode specific correlations among the gluon helicities; these correlations are transferred through hadronization and can ultimately be probed indirectly via final-state hadrons. A prominent example is the decay $V \to B\bar{B}$ (where $B$ denotes a baryon). Specifically, a key finding from Eqs.~\eqref{gamma one first}--\eqref{gamma four first} (or~\eqref{gamma one}--\eqref{gamma four}) is the strong dominance of decays from transversely polarized $V$ ($\lambda_V = \pm 1$) over those from longitudinally polarized $V$ ($\lambda_V = 0$). This pronounced transverse dominance implies the enhancement of transverse spin polarizations and correlations in the produced baryon pairs. Such effects can be probed experimentally via multi-dimensional angular-distribution fits and moment analyses of processes such as $V \to p\bar{p}$ or $\Lambda\bar{\Lambda}$~\cite{BESIII:2018cnd, BESIII:2025lzd}, thereby connecting our predictions for the polarized decay widths to measurable hadronic observables.

%%%%%%%%%%%%%%%%%%%%%%%%%%%%%%%%%%%%%%%%%%%%%%%%%%%%%%%%%%%%%%%%%%%%%%%%%%%%%%%%%%%%%%%%%%%%%%%%%%%%%%%%%%%%%%%%%%%%%%%%%%%%%%%%%
\subsection{Unpolarized decay widths}
\label{subsec:Decay widths}
%%%%%%%%%%%%%%%%%%%%%%%%%%%%%%%%%%%%%%%%%%%%%%%%%%%%%%%%%%%%%%%%%%%%%%%%%%%%%%%%%%%%%%%%%%%%%%%%%%%%%%%%%%%%%%%%%%%%%%%%%%%%%%%%%

The unpolarized decay width $\Gamma(V \to ggg)$ including $\hat{q}^2$-order relativistic corrections reads
\begin{eqnarray}
\Gamma (V  \to  ggg)= \frac{40(\pi^{2}-9)N_{V}^{2}\beta_{V}^{3}\alpha_{s}^{3}}{27\pi^{9/2} M }\left(1-\kappa\frac{\beta_{V}^{2}}{M^2} \right)
\end{eqnarray}
with $\kappa\equiv\frac{7(48+121\pi^2)}{48(\pi^2-9)}$. This expression shows that the relativistic correction, quantified by $\kappa \beta_V^{2}/M^{2}$, is significantly larger for $\psi(2S)$ (where $\kappa \beta_{\psi(2S)}^{2}/M^{2}\approx 2.70$) than for $\Upsilon(2S)$ (where $\kappa \beta_{\Upsilon(2S)}^{2} / M^{2} \approx 0.88$), owing to the lighter charm-quark mass. More importantly, for $\psi(2S)$, the decay width calculated to $\hat{q}^2$ order yields an unphysical negative value. This directly indicates the poor convergence of the relativistic expansion and underscores the necessity of including higher-order effects.

To address this issue and obtain a physically consistent prediction, we incorporate a selected subset of higher-order relativistic corrections through our phenomenological treatment. Summing the non-vanishing polarized widths over the helicity configurations of the four equivalence classes listed in the Appendix -- four configurations for each of $\Gamma_{1,2,3}$ and two for $\Gamma_{4}$ -- and averaging over the three polarizations of the initial vector meson, we reconstruct the unpolarized decay width from the polarized components as
\begin{eqnarray}\label{polarizedcombination}
\Gamma (V  \to  ggg)= \frac{1}{3} \left(4 \Gamma_1 + 4 \Gamma_2 + 4 \Gamma_3 + 2 \Gamma_4 \right),
\end{eqnarray}
where $\Gamma_1$-$\Gamma_4$ are given by Eqs.~(\ref{gamma one})--(\ref{gamma four}) and are calculated by incorporating $\hat{q}^2$-order relativistic corrections and systematically including a selected subset of higher-order effects, thereby restoring physical positivity. Their combination according to Eq.~(\ref{polarizedcombination}) provides a more reliable theoretical description of the decay width.

Since relativistic corrections arising from the internal motion of the bound quarks and QCD radiative corrections originating from additional gluon emissions during the hard process are governed by distinct dynamical scales, it is standard to treat them as factorizable at the present level of accuracy. Incorporating both types of corrections, we obtain the unpolarized decay width as the product of the relativistic correction factor and the QCD radiative correction factor:
\begin{eqnarray}\label{bothcorrections}
\Gamma (V  \to  ggg)= \frac{40(\pi^{2}-9)N_{V}^{2}\beta_{V}^{3}\alpha_{s}^{3}}{27\pi^{9/2} M }\left(1-\kappa\frac{\beta_{V}^{2}}{M^2} \right) \left( 1 - C_{V} \frac{\alpha_{s}}{\pi} \right).
\end{eqnarray}
Here, the relativistic correction factor $\left(1-\kappa\frac{\beta_{V}^{2}}{M^2} \right)$ includes only $\hat{q}^2$-order corrections, while the QCD radiative correction factor $\left( 1 - C_{V} \frac{\alpha_{s}}{\pi} \right)$ contains the $\mathcal{O}(\alpha_s)$ contribution. The process-dependent coefficient $C_{V}$ takes the values $C_{\psi(2S)} = 3.7$ and $C_{\Upsilon(2S)} = 4.9$, as determined in Ref.~\cite{Kwong:1987ak}.

However, as noted earlier, the $\hat{q}^2$-order treatment leads to unphysical results for $\psi(2S)$. To go beyond it, we include a phenomenological subset of higher-order relativistic effects. The improved expression is obtained by replacing the product of the leading-order width and the relativistic correction factor in Eq.~(\ref{bothcorrections}) with the combination of phenomenologically improved polarized widths:
\begin{eqnarray}
\Gamma (V  \to  ggg)= \frac{1}{3} \left(4 \Gamma_1 + 4 \Gamma_2 + 4 \Gamma_3 + 2 \Gamma_4 \right) \left( 1 - C_{V} \frac{\alpha_{s}}{\pi} \right),
\end{eqnarray}
where $\Gamma_1\text{--}\Gamma_4$ are given by Eqs.~(\ref{gamma one})\text{--}(\ref{gamma four}), and $C_{V}$ is understood as $C_{\psi(2S)}$ or $C_{\Upsilon(2S)}$ for the respective quarkonium.

We also compute the leptonic decay width $\Gamma(V \to e^+e^-)$. Including $\hat{q}^2$-order relativistic and one-loop QCD corrections, it is given by
\begin{eqnarray}
\Gamma (V  \to  e^{+}e^{-}) = \frac{12 e_{Q}^{2} \alpha^2 N_V^2 \beta_V^3}{ \pi^{7/2} M} \left(1 - \frac{28 \beta_{V}^{2}}{3M^{2}}\right) \left(1 - \frac{16 \alpha_s}{3\pi}\right),
\end{eqnarray}
where $e_Q$ denotes the electric charge of the heavy quark. Applying an analogous phenomenological treatment to that employed for the three-gluon decay yields the improved form
\begin{eqnarray}
\Gamma (V  \to  e^{+}e^{-}) = \frac{12 e_{Q}^{2} \alpha^2 N_V^2 \beta_V^3}{ \pi^{7/2} M} \biggl( \frac{27 \pi \bigl(a_0 (1+\frac{\beta_{V}^{2}}{M^2})-\frac{2 \sqrt{3}\beta_{V}}{9 \sqrt{\pi} M} (3+\frac{\beta_{V}^{2}}{M^{2}})^2\bigr)}{16 \left(\frac{\beta_{V}}{M}\right)^{10}}\biggr) \left(1 - \frac{16 \alpha_s}{3\pi}\right)
\end{eqnarray}
with $a_0=e^{3 M^2/(4 \beta_{V}^{2})} \operatorname{erfc}(\frac{\sqrt{3} M}{2 \beta_{V}} )$.

To substantially reduce the theoretical uncertainties, we construct the ratio of the hadronic to leptonic decay widths:
\begin{eqnarray}\label{Rv}
R_{V} = \frac{\Gamma(V  \to  ggg)}{\Gamma(V  \to  e^{+}e^{-})}.
\end{eqnarray}
In this ratio, the common nonperturbative factor $N_V^2\beta_V^3$, which is associated with the quarkonium wave function at the origin, cancels exactly. Consequently, $R_V$ provides a significantly cleaner observable, offering a more direct probe of the short-distance dynamics and thus yielding a more robust theoretical prediction.

For our numerical analysis, we take the meson masses and total widths from the PDG~\cite{ParticleDataGroup:2024cfk}. The QCD running coupling constants are taken as $\alpha_{s}(M_{\psi(2S)}/2)=0.31$ and $\alpha_{s}(M_{\Upsilon(2S)}/2)=0.21$, which are calculated through the two-loop renormalization group equation. We adopt the harmonic oscillator parameters $  \beta_{\psi(2S)} = 420\,\mathrm{MeV}  $ for $  \psi(2S)  $ and $  \beta_{\Upsilon(2S)} = 650\,\mathrm{MeV}  $ for $  \Upsilon(2S)  $. The value $  \beta_{\psi(2S)} = 420\,\mathrm{MeV}  $ is chosen to be consistent with the nodal position of the $  \psi(2S)  $ wave function extracted from lattice QCD~\cite{Nochi:2016wqg}, while $  \beta_{\Upsilon(2S)} = 650\,\mathrm{MeV}  $ lies within the range obtained in studies of highly excited bottomonium states~\cite{Weng:2018ebv}. The constants $  N_{\psi(2S)} = 0.232 \, \mathrm{MeV}^{-1/2}  $ for $  \psi(2S)  $ and $  N_{\Upsilon(2S)} = 0.167 \, \mathrm{MeV}^{-1/2}  $ for $  \Upsilon(2S)  $ are determined by fitting to the experimental branching ratios of the corresponding leptonic decays $  V \to e^+ e^-  $. With these inputs, we present our predictions for the branching ratios $\mathcal{B}(V \to ggg)$, $\mathcal{B}(V \to e^+e^-)$, and their ratio $R_V$ in Table~\ref{tab:predictions}. Theoretical uncertainties are omitted; they arise mainly from the scale choice for $\alpha_s$ and uncalculated higher-order effects, and are expected to be below $30\%$. The second column (``$\hat{q}^2$-order'') includes $\hat{q}^2$-order relativistic corrections, while the third (``Higher-order'') incorporates phenomenologically improved higher-order corrections; both include the known $\mathcal{O}(\alpha_s)$ QCD radiative corrections.

%%%%=============================================================
\begin{table}[!htbp]
  \caption{\label{tab:predictions} Theoretical predictions at $\hat{q}^2$-order relativistic correction and with higher-order relativistic corrections, together with experimental data, for the branching ratios $\mathcal{B}(V \to ggg)$, $\mathcal{B}(V \to e^{+}e^{-})$, and the ratio $R_V$ in $\psi(2S)$ and $\Upsilon(2S)$ decays.}
  \vspace{0.2cm}
  \centering
  \begin{tabular}{lccc}
  \hline\hline
  ~~&~~$\hat{q}^2$-order~~~ &~~Higher-order~~&~~Experiment~\cite{CLEO:2005mdr, CLEO:2009zjv, ParticleDataGroup:2024cfk}~~~  \\
  \hline
 $\mathcal{B}(\psi(2S)  \to  ggg)$~~&~~$ -88.8 \% $~~&~~$ 9.4 \% $~~&~~$(10.6 \pm 1.6 ) \%$ \\
 $\mathcal{B}(\psi(2S)  \to  e^{+}e^{-})$~~&~~$ 7.94 \times 10^{-3} $
 ~~&~~$ 8.04 \times 10^{-3} $~~&~~$(7.94 \pm 0.22) \times 10^{-3} $\\
 $R_{\psi(2S)}$                ~~&~~$ -111.8 $~~&~~$ 11.8 $ ~~&~~$ 13.4 \pm 2.4 $  \\
  \hline
 $\mathcal{B}(\Upsilon(2S)  \to  ggg)$~~&~~$ 13.8 \% $~~&~~$ 54.5 \% $~~&~~$(58.8 \pm 1.2) \% $ \\
 $\mathcal{B}(\Upsilon(2S)  \to  e^{+}e^{-})$~~&~~$ 1.91 \% $
 ~~&~~$ 1.91 \% $~~&~~$(1.91 \pm 0.16) \% $\\
 $R_{\Upsilon(2S)}$                ~~&~~$ 7.2 $~~&~~$ 28.5 $ ~~&~~$30.8 \pm 3.2 $ \\
  \hline\hline
  \end{tabular}
\end{table}
%%%%=============================================================

The numerical results in Table~\ref{tab:predictions} reveal several key features. First, as mentioned earlier, the predictions at $  \hat{q}^2  $-order relativistic correction (second column) yield unphysical negative values for both the branching ratio $  \mathcal{B}(\psi(2S) \to ggg)  $ and the ratio $  R_{\psi(2S)}  $. This dramatic failure arises because the three-gluon decay amplitude involves a convolution with multiple quark propagators, making it highly sensitive to the momentum distribution across the entire wave function. The characteristic nodal structure of the $2S$ radial wave function introduces significant destructive interference in this convolution, which is severely misrepresented when only the leading $  \hat{q}^2  $-order correction is included. In contrast, after applying our phenomenologically improved treatment, the results shift dramatically and come into good agreement with experimental data. This strongly indicates that higher-order relativistic corrections beyond $  \hat{q}^2  $ play a crucial role in the gluonic decay of the $2S$ state. A similar conclusion was previously reached in the exclusive hadronic decay $  \psi(2S) \to p\bar{p}  $~\cite{Kivel:2022qjy}.

Second, in the leptonic decay channel $  V \to e^+e^-  $, the predictions at $  \hat{q}^2  $-order (second column) are already very close to those with higher-order corrections (third column) and agree well with the experimental value. This suggests that the dominant relativistic corrections in the leptonic decay are largely captured already at the $  \hat{q}^2  $ level. The relatively mild impact of higher-order terms highlights the fast convergence of the relativistic expansion for this process and demonstrates its practical reliability for vector-meson leptonic widths.

Third, a striking contrast emerges when comparing the convergence behavior between the gluonic and leptonic decay channels. While the relativistic expansion converges rapidly for $  \Gamma(V \to e^+e^-)  $, it converges extremely slowly for $  \Gamma(V \to ggg)  $. This profound difference originates from the distinct dynamics of the two processes. The leptonic decay is mediated by a single virtual photon, and its width is predominantly sensitive to the wave function at the origin, where the nodal structure has a milder effect. In contrast, the amplitude for $  V \to ggg  $ involves a convolution with multiple quark propagators, making it acutely sensitive to the detailed momentum distribution across the entire wave function. The nodal structure introduces destructive interference in this convolution, which is severely misrepresented at $  \hat{q}^2  $ order. Thus, it is the combination of the nodal structure and the specific multi-gluon dynamics that leads to the dramatic failure of the low-order expansion in the gluonic channel.

Once both the dominant higher-order relativistic effects and the next-to-leading-order QCD radiative corrections are included, our predictions for $\mathcal{B}(V\to ggg)$, $\mathcal{B}(V\to e^{+}e^{-})$ and $R_{V}$ are in good agreement with the experimental data for both $\psi(2S)$ and $\Upsilon(2S)$.

%%%%%%%%%%%%%%%%%%%%%%%%%%%%%%%%%%%%%%%%%%%%%%%%%%%%%%%%%%%%%%%%%%%%%%%%%%%%%%%%%%%%%%%%%%%%%%%%%%%%%%%%%%%%%%%%%%%%%%%%%%%%%%%%%
\subsection{Harmonic oscillator parameter}
\label{subsec:Harmonic oscillator parameter}
%%%%%%%%%%%%%%%%%%%%%%%%%%%%%%%%%%%%%%%%%%%%%%%%%%%%%%%%%%%%%%%%%%%%%%%%%%%%%%%%%%%%%%%%%%%%%%%%%%%%%%%%%%%%%%%%%%%%%%%%%%%%%%%%%

As a cross-check, we further extract the harmonic oscillator parameter $\beta_V$ within our B-S framework. This parameter determines the characteristic momentum-space width of the quarkonium wave function. Using the ratio
\begin{eqnarray}
R_{V} = \frac{\Gamma(V  \to  ggg)}{\Gamma(V  \to  e^{+}e^{-})},
\end{eqnarray}
together with the experimental values~\cite{CLEO:2005mdr, CLEO:2009zjv, ParticleDataGroup:2024cfk}
\begin{eqnarray}
R_{\psi(2S)}^{exp}=13.4 \pm 2.4, \,\,\,\,\,\,\,\, R_{\Upsilon(2S)}^{exp}=30.8 \pm 3.2,
\end{eqnarray}
we obtain the following allowed ranges:
\begin{eqnarray}
\beta_{\psi(2S)} \in (360, 430)\, \mathrm{MeV}, \,\,\,\,\,\,\,\, \beta_{\Upsilon(2S)} \in (540, 670)\, \mathrm{MeV}.
\end{eqnarray}
%
%%%=============================================================
\begin{figure}[!!htb]
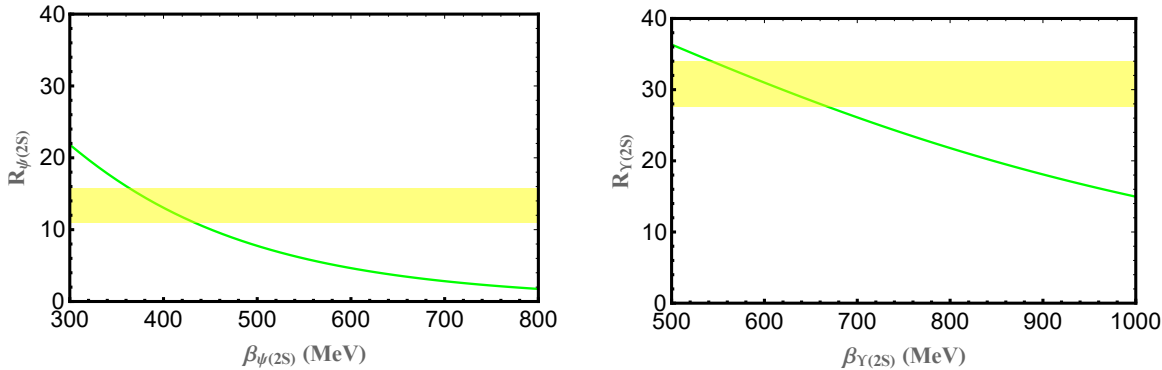

\centering
\includegraphics[width=0.43\textwidth]{figure1a.pdf}\hspace{0.5cm}
\includegraphics[width=0.43\textwidth]{figure1b.pdf}
\caption{\label{Rpu} Dependence of the ratio $R_{V}$ on the harmonic oscillator parameter $\beta_V$. The green curve shows our theoretical prediction. The yellow band represents the experimental value with its $1\sigma$ uncertainty. The intersection determines the extracted $\beta_V$ range.}
\end{figure}
%%%=============================================================

The functional dependence of the ratio $R_V$ on the harmonic oscillator parameter $\beta_V$ is displayed in Fig.~\ref{Rpu}. Our extracted $\beta_V$ values lie at the lower end of the ranges commonly reported in the literature for these states, with typical values being approximately $(400\text{--}700)\, \mathrm{MeV}$ for $\psi(2S)$ and $(600\text{--}1300)\, \mathrm{MeV}$ for $\Upsilon(2S)$~\cite{Peng:2012tr, Choi:2015ywa, Arifi:2022pal, Acharyya:2024tql, Godfrey:2015dia, Asghar:2023fvk, Gao:2025tob, Ahmad:2025hcr}. These broader ranges encompass values derived from various approaches, such as light-front quark models~\cite{Peng:2012tr, Choi:2015ywa, Arifi:2022pal, Acharyya:2024tql} and potential models~\cite{Godfrey:2015dia, Asghar:2023fvk, Gao:2025tob, Ahmad:2025hcr}. Our analysis, which employs the distinct observable $R_V$, provides an independent determination that refines these phenomenological estimates, consistently favoring the lower portion of the allowed intervals.

This preference for smaller $\beta_V$ values carries important physical implications. Because the B-S framework provides a fully relativistic treatment of bound-state dynamics, it yields a wave function that is more localized in momentum space than those from phenomenological models. This increased localization, which corresponds directly to a smaller $\beta_V$~\cite{Gao:2025tob}, consequently ensures a more reliable description of the high-momentum components that are often poorly captured by effective models. Hence, the larger $\beta_V$ values of many phenomenological models can be interpreted as an effective increase in the width parameter that compensates for their missing relativistic and high-momentum dynamics. Therefore, by calibrating this key parameter through processes that are acutely sensitive to such dynamics, our work establishes a concrete benchmark for non-perturbative models and enables a deeper understanding of heavy quarkonium structure.

%%%%%%%%%%%%%%%%%%%%%%%%%%%%%%%%%%%%%%%%%%%%%%%%%%%%%%%%%%%%%%%%%%%%%
\section{SUMMARY}
\label{sec:summary}

In this work, we have investigated the three-gluon decays of the radially excited vector quarkonia $  \psi(2S)  $ and $  \Upsilon(2S)  $ within the B-S formalism. By employing analytic harmonic oscillator wave functions that explicitly incorporate the nodal structure of the $2S$ states, we obtained fully analytic expressions for the decay amplitudes. To go beyond the $  \hat{q}^2  $-order relativistic corrections, we introduced a simple yet effective phenomenological treatment that incorporates the dominant higher-order contributions while preserving the correct low-momentum limit. Model-independent relations among polarized decay widths, derived from helicity-flip and phase-space symmetries, allow us to classify the non-vanishing contributions into four distinct groups, each with compact analytic formulas.

Our numerical results reveal a striking contrast between decay channels. At $  \hat{q}^2  $ order the gluonic width for $\psi(2S)$ and the ratio $  R_{\psi(2S)}  $ become unphysical and negative, exposing the breakdown of the low-order expansion caused by destructive interference from the $2S$ nodal structure in the multi-gluon convolution. The improved treatment restores positive, physically consistent predictions that agree well with experimental branching ratios for both $  \psi(2S)  $ and $  \Upsilon(2S)  $. In contrast, the leptonic decay $  V \to e^+e^-  $ already converges rapidly at $  \hat{q}^2  $ order, underscoring that the three-gluon channel is far more sensitive to the detailed momentum distribution across the entire wave function. As a cross-check, the harmonic oscillator parameter $  \beta_V  $ extracted from the ratio $  R_V  $ further confirms the reliability of our framework.

It is worth noting that, while our phenomenological treatment is not a first-principles derivation, its remarkably concise and physically transparent form offers a practical yet intuitive way to incorporate the dominant higher-order relativistic corrections beyond $  \hat{q}^2  $, together with the nodal structure characteristic of the $2S$ radial wave function. These findings establish gluonic decays of radially excited heavy quarkonia as a precision probe of relativistic bound-state dynamics, demonstrating that the internal nodal structure dramatically amplifies the sensitivity to higher-order effects in multi-gluon channels. This work thereby provides valuable insight into the unique role of excited-state internal structure in heavy quarkonium physics.

%%%%%%%%%%%%%%%%%%%%%%%%%%%%%%%%%%%%%%%%%%%%%%%%%%%%%%%%%%%%%%%%%%%%%
\section*{ACKNOWLEDGMENTS}
The work of C.-J. F. is supported by the Hubei Provincial Natural Science Foundation Youth Project under Grant No. 2024AFB151. The work of J.-K. H. is supported by the National Natural Science Foundation of China under Grant No. 12305086, the Open Fund of the Key Laboratory of Quark and Lepton Physics (MOE) under Grant No. QLPL2024P01, and the Project of Science and Technology Research Program of the Hubei Provincial Department of Education under Grant No. Q20222504.

%%%%%%%%%%%%%%%%%%%%%%%%%%%%%%%%%%%%%%%%%%%%%%%%%%%%%%%%%%%%%%%%%%%%%
\appendix
\section*{Appendix: Symmetry analysis and analytic $\hat{q}^{2}$-order expressions of the polarized decay widths}
\label{app:polarized}
\setcounter{equation}{0}
\renewcommand{\theequation}{A.\arabic{equation}}
%%%%%%%%%%%%%%%%%%%%%%%%%%%%%%%%%%%%%%%%%%%%%%%%%%%%%%%%%%%%%%%%%%%%%

For completeness we collect here the symmetry analysis of the polarized decay widths $\Gamma^{ggg}_{[\lambda_V,\lambda_1,\lambda_2,\lambda_3]}$ and the analytic $\hat{q}^{2}$-order expressions for the four equivalence classes $\Gamma_1$--$\Gamma_4$. The derivations follow exactly the methodology established in our previous work on the ground-state decays $J/\psi\to ggg$ and $\Upsilon(1S)\to ggg$~\cite{Jiang:2025cks}, now with the Dirac structures extended to ${\cal O}(\hat{q}^{2})$ and the $1S$ scalar radial function replaced by the $2S$ one in Eq.~(\ref{fhatq}).

Owing to helicity-flip symmetry and phase-space symmetry, the following relations are satisfied:
\begin{eqnarray}
\Gamma_{[\lambda_V,\lambda_1,\lambda_2,\lambda_3]}^{ggg} = \Gamma_{[-\lambda_V,-\lambda_1,-\lambda_2,-\lambda_3]}^{ggg}, \quad  \quad
\Gamma_{[\lambda_V,\lambda_1,\lambda_2,\lambda_3]}^{ggg} = \Gamma_{[\lambda_V,\lambda_1,\lambda_3,\lambda_2]}^{ggg}.
\end{eqnarray}
Furthermore, after summing over the polarizations $\lambda_V$ of the initial quarkonium to eliminate the preferred direction defined by the initial polarization, the full permutation symmetry among the three identical gluons is restored, leading to the additional equalities:
\begin{eqnarray}
\sum_{\mathrm{\lambda_V=-1,0,1}}\Gamma_{[\lambda_V,1,1,-1]}^{ggg}
&=&\sum_{\mathrm{\lambda_V=-1,0,1}}\Gamma_{[\lambda_V,1,-1,1]}^{ggg}
=\sum_{\mathrm{\lambda_V=-1,0,1}}\Gamma_{[\lambda_V,-1,1,1]}^{ggg}\nonumber \\
&=&\sum_{\mathrm{\lambda_V=-1,0,1}}\Gamma_{[\lambda_V,-1,1,-1]}^{ggg}
=\sum_{\mathrm{\lambda_V=-1,0,1}}\Gamma_{[\lambda_V,1,-1,-1]}^{ggg}
=\sum_{\mathrm{\lambda_V=-1,0,1}}\Gamma_{[\lambda_V,-1,-1,1]}^{ggg}, \nonumber \\
\sum_{\mathrm{\lambda_V=-1,0,1}}\Gamma_{[\lambda_V,1,1,1]}^{ggg}
&=&\sum_{\mathrm{\lambda_V=-1,0,1}}\Gamma_{[\lambda_V,-1,-1,-1]}^{ggg}.
\end{eqnarray}
These relations originate directly from the fundamental symmetries of QCD and the bosonic nature of the gluons and provide a model-independent self-consistency check for the present calculation.

The following helicity configurations yield identically zero width:
\begin{eqnarray}
\Gamma_{[1, 1, 1, 1]}^{ggg} =&& \Gamma_{[-1, -1, -1, -1]}^{ggg} = 0, \quad \quad \Gamma_{[1, -1, -1, -1]}^{ggg} = \Gamma_{[-1, 1, 1, 1]}^{ggg} = 0, \quad \quad \Gamma_{[0, 1, 1, 1]}^{ggg} = \Gamma_{[0, -1, -1, -1]}^{ggg} = 0,\nonumber \\
&&\Gamma_{[1, -1, 1, 1]}^{ggg} = \Gamma_{[-1, 1, -1, -1]}^{ggg} = 0, \quad \quad \Gamma_{[0, 1, -1, -1]}^{ggg} = \Gamma_{[0, -1, 1, 1]}^{ggg} = 0.
\end{eqnarray}
The remaining non-vanishing polarized decay widths fall into four equivalence classes labeled $\Gamma_1$, $\Gamma_2$, $\Gamma_3$ and $\Gamma_4$:
\begin{eqnarray}\label{group one}
\Gamma_{[1, 1, 1, -1]}^{ggg} & = & \Gamma_{[-1, -1, -1, 1]}^{ggg} = \Gamma_{[-1, -1, 1, -1]}^{ggg} = \Gamma_{[1, 1, -1, 1]}^{ggg} = \Gamma_1,
\end{eqnarray}
\begin{eqnarray}\label{group two}
\Gamma_{[1, -1, 1, -1]}^{ggg} & = & \Gamma_{[-1, 1, -1, 1]}^{ggg} = \Gamma_{[-1, 1, 1, -1]}^{ggg} = \Gamma_{[1, -1, -1, 1]}^{ggg}  = \Gamma_2,
\end{eqnarray}
\begin{eqnarray}\label{group three}
\Gamma_{[0, 1, 1, -1]}^{ggg} & = & \Gamma_{[0, -1, -1, 1]}^{ggg} = \Gamma_{[0, -1, 1, -1]}^{ggg} = \Gamma_{[0, 1, -1, 1]}^{ggg}  = \Gamma_3,
\end{eqnarray}
\begin{eqnarray}\label{group four}
\Gamma_{[1, 1, -1, -1]}^{ggg} & = & \Gamma_{[-1, -1, 1, 1]}^{ggg}  = \Gamma_4,
\end{eqnarray}
where $\Gamma_{1,2,3}$ each collect four equivalent helicity configurations, while $\Gamma_{4}$ collects only two. Performing the analytic integration over the internal momentum of the heavy quarkonium and the three-body phase space, the four classes admit the closed-form $\hat{q}^{2}$-order expressions
\begin{eqnarray}\label{gamma one first}
\Gamma_1 & = & \frac{5 \alpha_s^3 N_V^2 \beta_V^3 (3 \pi^2 - 16)}{216 M \pi^{9/2}} \left(1 - \frac{7 (1264 + 111 \pi^2)}{6 (3 \pi^2 - 16)} \frac{\beta_V^2}{M^2}\right),
\end{eqnarray}
\begin{eqnarray}\label{gamma two first}
\Gamma_2 & = & \frac{5 \alpha_s^3 N_V^2 \beta_V^3 (29 \pi^2 - 280)}{216 M \pi^{9/2}} \left(1 - \frac{7 (265 \pi^2 - 1424)}{6 (29 \pi^2 - 280)} \frac{\beta_V^2}{M^2}\right),
\end{eqnarray}
\begin{eqnarray}\label{gamma three first}
\Gamma_3 & = & \frac{5 \alpha_s^3 N_V^2 \beta_V^3}{27 M \pi^{9/2}} \left(1 - \frac{7 (88 + 27 \pi^2)}{12} \frac{\beta_V^2}{M^2}\right),
\end{eqnarray}
\begin{eqnarray}\label{gamma four first}
\Gamma_4 & = & \frac{20 \alpha_s^3 N_V^2 \beta_V^3 (\pi^2 - 9)}{27 M \pi^{9/2}} \left(1 - \frac{7 (48 + 121 \pi^2)}{48 (\pi^2 - 9)} \frac{\beta_V^2}{M^2}\right).
\end{eqnarray}
For $\psi(2S)$ these $\hat{q}^{2}$-order expressions exhibit the slow convergence of the relativistic expansion discussed in the main text; the phenomenologically improved expressions used in the main analysis are given in Eqs.~(\ref{gamma one})--(\ref{gamma four}).

\newpage

%%====================================================================
%%%%%%%%%%%%%%%%%%%%%%%%%%    References    %%%%%%%%%%%%%%%%%%%%%%%%%%
%%%%==================================================================
\providecommand{\href}[2]{#2}\begingroup\raggedright\endgroup

\end{document}